\documentstyle[12pt]{article}
\begin{document}
\begin{center}
CHAOS SYNCHRONIZATION IN THE MULTI-FEEDBACK IKEDA MODEL\\
E.M.Shahverdiev \\
Institute of Physics, 33,H.Javid Avenue, Baku-370143, Azerbaijan\\
ABSTRACT\\
\end{center}
We investigate synchronization between two unidirectionally 
coupled chaotic multi-feedback Ikeda systems and 
find both the existence and stability conditions for anticipating, lag, and 
complete synchronizations. Generalization of the approach to a wide 
class of nonlinear systems is also presented.\\
PACS number(s):05.45.Xt, 05.45.Vx, 42.55.Px, 42.65.Sf\\
\begin{center}
1.INTRODUCTION
\end{center}
\indent Seminal papers on chaos synchronization [1] have stimulated a wide range of 
research activity [2]. Synchronization phenomena in coupled systems have been especially 
extensively studied in the context of laser dynamics, electronic circuits, chemical and 
biological systems [2]. Application of chaos synchronization can be found in secure 
communication, optimization of nonlinear system performance, modeling brain activity and 
pattern recognition phenomena [2].\\
\indent Due to finite signal transmission times, switching speeds and memory effects 
systems with both a single and  multiple delays are ubiquitous in nature and technology [3]. 
Dynamics of multi-feedback systems are representative examples of the multi-delay systems. 
Therefore the study of synchronization phenomena in time-delayed systems is of high 
practical 
importance.Prominent examples of such dynamics can be found in biological and biomedical 
systems,
laser physics,integrated communications [3]. In laser physics such a situation arises in 
lasers subject 
to two or more optical or elctro-optical feedback. Second optical feedback could be 
useful to stabilize laser intensity [4]. Chaotic behaviour of laser systems with two 
optical feedback mechanism is studied in recent works [5]. To the best of our knowledge 
chaos synchronization between the multi-feedback systems is to be investigated yet. Having 
in mind enormous application implications of chaos synchronization e.g. in secure 
communication, investigation of synchronization regimes in the multi-feedback systems is 
of certain importance.\\
\indent  In this paper {\it for the first time} we investigate synchronization between 
two unidirectionally coupled chaotic multi-feedback Ikeda systems and find both the 
existence and stability conditions for different synchronization 
regimes. We also present generalization of the approach to the wide class of nonlinear 
systems.\\
\begin{center}
2.SYNCHRONIZATION BETWEEN THE MULTI-FEEDBACK IKEDA SYSTEMS\\
\end{center}
\indent Consider synchronization between the multi-feedback Ikeda systems,
$$\hspace*{5cm}\frac{dx}{dt}=-\alpha x + m_{1} \sin x_{\tau_{1}}
+m_{2} \sin x_{\tau_{2}},\hspace*{4.6cm}(1)$$
$$\hspace*{5cm}\frac{dy}{dt}=-\alpha y + m_{3} \sin y_{\tau_{1}}
+ m_{4} \sin y_{\tau_{2}} + K \sin x_{\tau_{3}},\hspace*{3.5cm}(2)$$
with positive $\alpha_{1,2}$ and $-m_{1,2,3,4}$. Throughout this 
paper $x_{\tau}\equiv x(t-\tau)$.\\ 
\indent  This investigation is of considerable practical importance, as the equations of 
the class B lasers with feedback (typical representatives of class B are solid-state, 
semiconductor, and low pressure $CO_{2}$ lasers [6]) can be reduced to an equation of the 
Ikeda type [7].\\ 
The Ikeda model was introduced to describe the dynamics of an optical bistable resonator, 
plays an important role in electronics and physiological studies and is well-known for 
delay-induced chaotic behavior [8-9], see also e.g. [10].
Physically $x$ is the phase 
lag of the electric field across the resonator; $\alpha$ is the relaxation coefficient 
for the driving $x$ and driven $y$ dynamical variables;$m_{1,2}$ and $m_{3,4}$ are the 
laser intensities injected into the driving and driven systems,respectively. $\tau_{1,2}$ 
are the feedback delay times in the coupled systems;$\tau_{3}$ is the coupling delay time 
between systems $x$ and $y$;$K$ is the coupling rate between the driver $x$ and the 
response system $y$.\\
We find that systems (1) and (2) can be synchronized on the synchronization 
manifold
$$\hspace*{7cm}y=x_{\tau_{3}-\tau_{1}}.\hspace*{8cm}(3)$$
as the error signal $\Delta=x_{\tau_{3}-\tau_{1}}-y$  for small $\Delta$ 
under the condition
$$\hspace*{6cm}m_{1}-K=m_{3}, m_{2}=m_{4}\hspace*{6.5cm}(4)$$
obey the following dynamics  
$$\hspace*{5cm}\frac{d\Delta}{dt}= -\alpha\Delta + m_{3} \Delta_{\tau_{1}} \cos x_{\tau_{3}} + m_{2} \Delta_{\tau_{2}} \cos  x_{\tau_{2}+\tau_{3}-\tau_{1}}.\hspace*{2.8cm}(5)$$
It is obvious that $\Delta= 0$ is a solution of system (5). We notice that 
for $\tau_{3}>\tau_{1}$,$\tau_{3}=\tau_{1}$, and $\tau_{3}<\tau_{1}$ 
 (3) is the retarded, complete, and anticipating synchronization manifold [10], 
respectively.\\
To study the stability of the synchronization manifold $y=x_{\tau_{3}-\tau_{1}}$ one 
can use the Krasovskii-Lyapunov 
functional approach. According to [3], the sufficient stability condition for the trivial 
solution $\Delta=0$ of time-delayed 
equation $\frac{d\Delta}{dt}=-r(t)\Delta + s_{1}(t)\Delta_{\tau_{1}}+ s_{2}(t)\Delta_{\tau_{2}} $ 
is: $r(t)>\vert s_{1}(t) \vert + \vert s_{2}(t) \vert $.\\
Thus we obtain that the sufficient stability condition for the synchronization 
manifold $y=x_{\tau_{3}-\tau_{1}}$ (3) can be written as:
$$\hspace*{5cm}\alpha > \vert m_{3} \vert +\vert m_{2} \vert.\hspace*{9cm}(6)$$ 
As Eq.(5) is valid for small $\Delta$ stability condition (6) found above holds locally.
Conditions (4) are the existence conditions for the synchronization manifold (3) 
between unidirectionally coupled multi-feedback systems (1) and (2).\\
Analogously we find that $y=x_{\tau_{3}-\tau_{2}}$ is the synchronization manifold 
between systems (1) and (2) with the existence $m_{2}-K=m_{4}$ 
and $m_{1}=m_{3}$ and stability 
conditions $\alpha > \vert m_{3} \vert + \vert m_{4} \vert $.\\
\indent One can generalize the previous rezults to $n$-tuple feedback Ikeda 
systems. Applying the error dynamics approach to synchronization between the following 
Ikeda models
$$\hspace*{1cm}\frac{dx}{dt}=-\alpha x +m_{1x} \sin x_{\tau_{1}} 
+ m_{2x} \sin x_{\tau_{2}}+ \cdots +m_{nx}\sin x_{\tau_{n}},\hspace*{5.3cm}(7)$$
$$\hspace*{2cm}\frac{dy}{dt}=-\alpha y +m_{1y} \sin y_{\tau_{1}}
+m_{2y} \sin y_{\tau_{2}} +m_{ny} \sin y_{\tau_{n}}+ k \sin x_{\tau_{k}},\hspace*{3.5cm}(8)$$
we find that the existence and sufficent stability conditions e.g. for the 
synchronization manifold $y=x_{\tau_{k}-\tau_{1}}$ are:$m_{1x}-k=m_{1y},m_{nx}=m_{ny}$ 
and $\alpha > \vert m_{1y} \vert + \vert m_{2y}\vert + \cdots + \vert m_{ny} \vert $,
respectively. For the synchronization manifold  $y=x_{\tau_{k}-\tau_{2}}$, $m_{2x}-k=m_{2y}$ 
and $m_{nx}=m_{ny}$ are the existence conditions, 
and  $\alpha >\vert m_{1y} \vert +\vert m_{2y} \vert + \cdots + \vert m_{ny} \vert$ is 
the sufficient stability condition.\\
\begin{center}
3.GENERAL APPROACH\\
\end{center}
\indent Consider synchronization between the double-feedback systems of general form ,
$$\hspace*{5cm}\frac{dx}{dt}=-\alpha x + m_{1} f( x_{\tau_{1}})
+m_{2} f( x_{\tau_{2}}),\hspace*{5.3cm}(9)$$
$$\hspace*{5cm}\frac{dy}{dt}=-\alpha y + m_{3} f( y_{\tau_{1}})
+ m_{4} f (y_{\tau_{2}}) + K f (x_{\tau_{3}}),\hspace*{3.5cm}(10)$$
where $f$ is differentiable generic nonlinear function.\\ 
One finds that under the condition 
$$\hspace*{6cm}m_{1}-K=m_{3}, m_{2}=m_{4}\hspace*{6.5cm}(11)$$
Eqs. (9) and (10) admit the synchronization manifold 
$$\hspace*{7cm}y=x_{\tau_{3}-\tau_{1}}.\hspace*{7.7cm}(12)$$
This follows from the dynamics of the error $\Delta=x_{\tau_{3}-\tau_{1}}-y$ 
$$\hspace*{5cm}\frac{d\Delta}{dt}= -\alpha\Delta + m_{3} \Delta_{\tau_{1}} 
f'( x_{\tau_{3}}) + m_{2} \Delta_{\tau_{2}} f' (x_{\tau_{2}+\tau_{3}-\tau_{1}}).\hspace*{2.7cm}(13)$$
Here $f'$ stands for the derivative of $f$ with respect to time.
The sufficient stability condition of the trivial solition $\Delta=0$ of (13) can be found 
from  the Krasovskii-Lyapunov functional approach [3]: 
$$\hspace*{5cm}\alpha > \vert m_{3}f'( x_{\tau_{3}})\vert +\vert m_{2} f' (x_{\tau_{2}+\tau_{3}-\tau_{1}}) \vert.\hspace*{5.3cm}(14)$$ 
Analogously one finds both the existence and ssufficient stability conditions for 
synchronization  manifold $y=x_{\tau_{3}-\tau_{2}}$. Further generalization of the approach to 
$n$-tuple feedback systems of type (9) and (10) is straightforward. 
\begin{center}
4.CONCLUSIONS\\
\end{center}
\indent 
{\it For the first time} we have investigated synchronization between two 
unidirectionally coupled chaotic multi-feedback Ikeda systems and 
find both the existence and stability conditions for anticipating, lag, and 
complete synchronization regimes. Generalization of the approach to the wide class of 
chaotic systems is also presented.\\
Having in mind different application possibilities of chaos synchronization, 
synchronization in multi-feedback systems can provide more flexibility e.g. in obtaining 
different anticipating time scales, etc. and opportunities in practical applications. \\
It is well known that laser arrays hold great promise for space communication 
applications, which require compact sources with high optical intensities.The most 
efficient result can be achieved when the array elements are synchronized. Additional 
feedback mechanizm could be useful to stabilize nonlinear system's output, e.g. 
laser intensity.\\
~\\
5.ACKNOWLEDGEMENTS\\
~\\
The author kindly acknowledges support from the Abdus Salam ICTP (Trieste, 
Italy) Associate scheme.\\

\end{document}